\documentclass[a4paper,fleqn]{cas-dc}
\usepackage[numbers, sort&compress]{natbib}
\def\tsc#1{\csdef{#1}{\textsc{\lowercase{#1}}\xspace}}
\tsc{WGM}
\tsc{QE}

\usepackage[utf8]{inputenc}
\usepackage{adjustbox}
\usepackage{amsmath,amssymb,amsfonts}
\usepackage{algorithmic}
\usepackage[]{algorithm2e}
\usepackage{adjustbox}
\usepackage{aeguill}
\usepackage{amssymb}
\usepackage{balance}
\usepackage[justification=centering,font={small,it},labelfont={small,bf},tableposition=top]{caption}
\usepackage{subcaption}
\usepackage{booktabs}
\usepackage{color}
\usepackage{commath}
\usepackage{epstopdf}
\usepackage{float}
\usepackage{graphicx}
\usepackage{indentfirst}
\usepackage{longtable}
\usepackage{lscape}
\usepackage{mathrsfs}
\usepackage{multicol, blindtext}
\usepackage{multirow}
\usepackage{natbib}
\usepackage{placeins}
\usepackage{scalefnt}
\usepackage{supertabular}
\usepackage{tabu}
\usepackage{tabularx}
\usepackage{textcomp}
\usepackage{url}
\usepackage{verbatim}
\usepackage{xcolor}
\usepackage[normalem]{ulem}
\usepackage{algorithmic}
\usepackage{graphicx}
\usepackage{textcomp}
\usepackage{subcaption}
\usepackage{xcolor}
\usepackage{float}
\usepackage{url}
\usepackage{hyperref}
\usepackage{ulem}
\newcommand{\blue}[1]{\textcolor{black}{#1}}

\usepackage{lscape}
\usepackage{pifont}
\usepackage{xcolor}
\newcommand{\cmark}{\ding{51}}%
\newcommand{\xmark}{\ding{55}}%

\newcommand\cmarkfillcircled[1][cyan]{%
    \ooalign{%
        \hidewidth
        \kern0.65ex\raisebox{-0.9ex}{\scalebox{3}{\textcolor{#1}{\textbullet}}}
        \hidewidth\cr
        \cmark\cr
    }%
}
\newcommand\xmarkfillcircled[1][red]{%
    \ooalign{%
        \hidewidth
        \kern0.65ex\raisebox{-0.9ex}{\scalebox{3}{\textcolor{#1}{\textbullet}}}
        \hidewidth\cr
        \xmark\cr
    }%
}

\makeatletter
\renewcommand\section{\@startsection{section}{1}{\z@}%
    {5pt plus 2pt minus 2pt}%
    {1pt plus 1pt minus 1pt}%
    {\normalfont\large\bfseries}}
\renewcommand\subsection{\@startsection{subsection}{2}{\z@}%
    {3pt plus 2pt minus 2pt}%
    {1pt plus 1pt minus 1pt}%
    {\normalfont\bfseries}}

\providecommand{\doi}{--}

\providecommand{\AnhNTAuthor}{Tuan Anh Nguyen}
\newcommand{\AnhNTAuthorOCRID}{0000-0002-9123-3584}

\providecommand{\DugkiMinAuthor}{Dugki Min}

\providecommand{\EunmiChoiAuthor}{Eunmi Choi}

\providecommand{\JaeWooLeeAuthor}{Jae-Woo Lee}

\providecommand{\AckCarlos}{This work was carried out with financial support from the Fundação Coordenação de Aperfeiçoamento de Pessoal de Nível Superior (CAPES) under process PDPG-POSDOC-AUXPE No. 88881.830176/2023-01.}

\providecommand{\AckProfessorLeeUAM}{This research was partially supported by Basic Science Research Program through the National Research Foundation of Korea (NRF) funded by the Ministry of Education (No. 2020R1A6A1A03046811).}

\providecommand{\AckProfessorMinMidsizedResearcher}{This research was supported by the Basic Science Research Program through the National Research Foundation of Korea (NRF) funded by the Ministry of Education (2021R1A2C2094943)}

\begin{document}
    \let\WriteBookmarks\relax
    \def\floatpagepagefraction{1}
    \def\textpagefraction{.001}
    
    \title [mode = title]{Transactional Dynamics in Hyperledger Fabric: A Stochastic Modeling and Performance Evaluation of Permissioned Blockchains}
    \author[1]{Carlos Melo}
    \ead{carlos.alexandre@ufpi.edu.br}
    \affiliation[1]{Laboratory of Applied Research to Distributed Systems (PASID), Federal University of Piaui (UFPI), Picos, Piaui 64607-670, Brazil}
    \author[1]{Glauber Gonçalves}
    \ead{ggoncalves@ufpi.edu.br}
    \author[1]{Francisco Airton Silva}
    \ead{faps@ufpi.edu.br}
    \author[1]{Iure Fé}
    \ead{iure.fe@ufpi.edu.br}
    \author[1]{Ericksulino Moura}
    \ead{ericksulino@ufpi.edu.br}
    \author[1]{André Soares}
    \ead{andre.soares@ufpi.edu.br}
    
     \author[2]{\EunmiChoiAuthor}
     \ead{emchoi@kookmin.ac.kr}
     \affiliation[2]{School of Software, College of Computer Science, Kookmin University, Seoul 02707, South Korea}  
     
     \author[3,4]{\DugkiMinAuthor}
     \ead{dkmin@konkuk.ac.kr}
     \affiliation[3]{Department of Computer Science and Engineering, College of Engineering, Konkuk University, Seoul 05029, South Korea}
     \affiliation[4]{Department of Artificial Intelligence, Graduate School, Konkuk University, Seoul 05029, South Korea}
     
     \author[5]{\JaeWooLeeAuthor}
     \ead{jwlee@konkuk.ac.kr}
     \affiliation[5]{Department of Mechanical and Aerospace Engineering, Konkuk University, Seoul 05029, South Korea}     
               
     \author[3,6]{\AnhNTAuthor}
     [
     orcid=\AnhNTAuthorOCRID
     ]
     \ead{anhnt2407@konkuk.ac.kr}
     \affiliation[2]{Konkuk Aerospace Design-Airworthiness Institute (KADA), Konkuk University, Seoul 05029, South Korea}
     
     {\footnotesize \cortext[1]{Corresponding authors: \{anhnt2407, dkmin, jwlee\} @konkuk.ac.kr}}
     
    \nonumnote{\AckCarlos; \AckProfessorLeeUAM; \AckProfessorMinMidsizedResearcher }
    
    \shorttitle{Optimizing Hyperledger Fabric: A Stochastic Petri Net Approach}
    \shortauthors{Carlos Melo; \textit{et. al.} 2024}

\begin{abstract}
   Blockchain, often integrated with distributed systems and security enhancements, has significant potential in various industries. However, environmental concerns and the efficiency of consortia-controlled permissioned networks remain critical issues. We use a Stochastic Petri Net model to analyze transaction flows in Hyperledger Fabric networks, achieving a 95\% confidence interval for response times. This model enables administrators to assess the impact of system changes on resource utilization. Sensitivity analysis reveals major factors influencing response times and throughput. Our case studies demonstrate that block size can alter throughput and response times by up to 200\%, underscoring the need for performance optimization with resource efficiency.
\end{abstract}



\maketitle

\section{Introduction}
\label{sec:intro}


\textcolor{black}{
    Hyperledger Fabric (HLF) \cite{Androulaki2018}, a permissioned blockchain by the Hyperledger Foundation, is designed for industrial applications requiring node identification. 
    Transactions are processed through endorsement, ordering, and committing, integrating validated transactions into the blockchain \cite{Melo2022}. 
    Unlike public blockchains, HLF mandates participant identification, enhancing security and privacy. 
    Its modular architecture allows customization of components, such as consensus mechanisms and membership services, to meet organizational needs. 
    The execute-order-validate paradigm improves scalability and performance through parallel transaction execution and validation. 
    HLF supports private data collections and channels for confidential transactions, and its query capabilities via CouchDB or LevelDB accommodate diverse applications. 
    This paper uses a Stochastic Petri Net (SPN) model to analyze HLF's transactional dynamics, focusing on mean response time (MRT), throughput, and resource utilization. 
    We provide sensitivity analysis for optimizing configurations and improving system efficiency. 
    Current research on HLF lacks validated models and formal bottleneck detection methods, focusing on older versions. 
    Our study addresses these gaps with a robust SPN, comprehensive sensitivity analysis, and experimental validation for accurate performance optimization.
    }

The main \textbf{contributions} of this paper are as follows: \textit{(i.) Stochastic Petri Net (SPN) model for Hyperledger Fabric:} Introduces a specialized SPN model to analyze transaction flows in permissioned blockchain environments, particularly in Hyperledger Fabric networks. \textit{(ii.) Experimental validation of SPN model:} Conducts thorough experiments to validate the SPN model, demonstrating its accuracy and reliability in simulating real-world blockchain operations. \textit{(iii.) Sensitivity analysis of system parameters:} Provides a comprehensive sensitivity analysis to explore the influence of variations in system parameters on key performance metrics. And the key \textbf{findings} of this study are: \textit{(i.) Performance implications of block size:} Uncovers that block size has a substantial impact on system performance, notably affecting throughput by up to 200\% and mean response time. \textit{(ii.) Critical system parameters affecting efficiency:} Identifies transaction arrival rate and resource allocation as pivotal factors influencing overall system efficiency and responsiveness.

\textcolor{black}{This work addresses limitations in Hyperledger Fabric (HLF) research, such as focusing on earlier versions and lacking validated models and formal bottleneck detection. Previous studies inadequately evaluated key performance metrics like throughput, latency, and discard rate. We introduce a robust Stochastic Petri Net (SPN) model for HLF 2.5+, validated through extensive experiments, and provide comprehensive sensitivity analysis. This approach bridges research gaps and offers actionable insights for optimizing HLF configurations.}

\textcolor{black}{Our study provides practical applications for optimizing Hyperledger Fabric (HLF) networks, which are crucial in industries like finance, supply chain, healthcare, and government. By utilizing our Stochastic Petri Net (SPN) model, system administrators can fine-tune HLF configurations, identify bottlenecks, plan for scalability, and optimize resource utilization. The model's detailed sensitivity analysis offers actionable insights into performance tuning and resource management. Our experimental validation ensures that these findings are applicable to real-world HLF deployments, enhancing the reliability and practical relevance of our recommendations.}

This paper is organized as follows: Section \ref{sec:related} reviews blockchain modeling and performance; Section \ref{sec:proposed} introduces our model; Section \ref{sec:measurement} details experiments and validation; Section \ref{sec:results} discusses findings; Section \ref{sec:conclusion} concludes.

\section{Related Works}
\label{sec:related}

Although the Hyperledger Fabric (HLF) platform's performance has been widely analyzed in earlier research, this paper aims to address some unexplored aspects.

\blue{Melo et al.~\cite{Melo2022} presented models to assess resource utilization on the HLF platform using Continuous Time Markov Chains (CTMC) and Stochastic Petri Nets (SPNs). Their findings highlighted the utility of these models in planning HLF applications, mainly through sensitivity analysis for detecting infrastructure bottlenecks. However, their research overlooked vital performance metrics such as throughput, latency, and discard rate, which this paper aims to evaluate comprehensively.}

\blue{Jiang et al.~\cite{Jiang2020} introduced a hierarchical model for transaction processes in Hyperledger Fabric v1.4, which considered the constraints of endorsement and block timeouts. This model enabled the calculation of performance metrics such as platform throughput, transaction discard probability, and mean response time.}

\blue{Wu et al.~\cite{wu_acm_ease2022} utilized queuing theory and a two-dimensional continuous-time Markov process to analyze HLF 2.0, focusing on performance metrics including throughput and resource utilization. This approach provided valuable insights but did not incorporate a formal bottleneck detection method.}

\blue{Ke and Park~\cite{Ke2023} created queuing models to evaluate HLF performance, particularly analyzing queue length and mean response time under different service rates for endorsing and validation. Their work highlighted the impact of service rates on system performance but also lacked a formal bottleneck detection framework.}

\blue{Yuan et al.~\cite{yuan2020performance} employed Generalized Stochastic Petri Nets (GSPN) to investigate the effects of arrival rates, block size, and timeout settings on HLF throughput and latency. Their study provided a detailed analysis of these parameters but did not address formal bottleneck detection.}

\blue{Sukhwani et al.~\cite{Sukhwani2018} used Stochastic Reward Networks (SRN) to evaluate performance metrics and estimate average queue sizes throughout the HLF transaction process. Although this research involved experimental validation, it did not include a formal methodology for bottleneck detection, and its focus on earlier HLF versions may limit applicability to current real-world conditions.}

This paper builds on these previous efforts by introducing a robust methodology for bottleneck detection within the HLF 2.5+ version. Our approach integrates formal model validation with percentage difference sensitivity analysis to pinpoint potential bottlenecks in throughput, mean response time, and the impact of queue size parameters on resource utilization.

\begin{figure*}[htpb]
    \centering    
    \begin{subfigure}{\textwidth}
        \centering
        \includegraphics[width=0.75\linewidth]{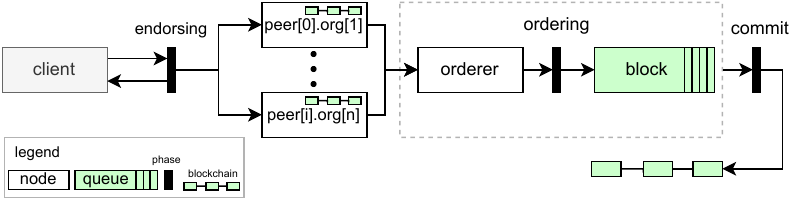}
        \caption{Transaction Flow}
        \label{fig_overview}
    \end{subfigure}%
    \hfill
    \begin{subfigure}{\textwidth}
        \centering
        \includegraphics[width=0.65\linewidth]{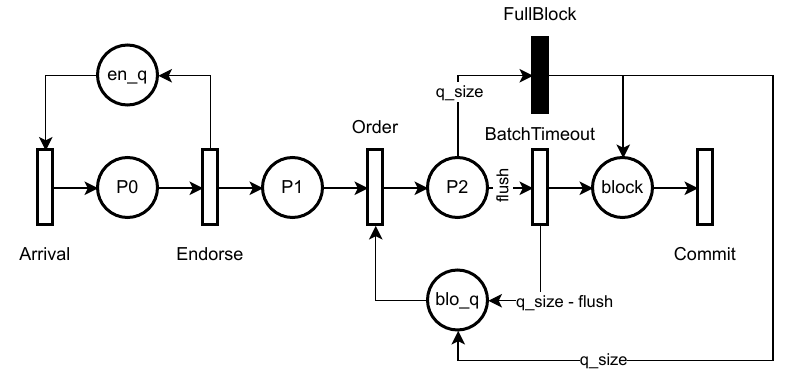}
        \caption{Performance Model}
        \label{fig_spn_model}
    \end{subfigure}
    \vspace{-10pt}
    \caption{Hyperledger Fabric: Architecture and Stochastic Model}
    \label{fig_combined_figs}
\end{figure*}

\section{Architecture and Proposed Model}
\label{sec:proposed}

\subsection{Hyperledger Fabric Architecture}

\blue{Hyperledger Fabric (HLF) is an open-source permissioned blockchain~\cite{Androulaki2018}. Unlike public blockchains like Bitcoin and Ethereum, which rely on anonymous participation, HLF focuses on networks among entities with common goals but potential trust issues due to business competition and requires participant identification within private communication channels. HLF features identifiable nodes, varying security levels for transaction execution with endorsement policies, and an architecture that executes-orders-validates transactions, outperforming Bitcoin and Ethereum's order-execute model while enhancing security~\cite{thakkar2018performance}.}

Figure \ref{fig_overview} outlines the transaction flow in the HLF network, employing a three-step process: \textit{endorsement}, \textit{ordering}, and \textit{commit}. \textcolor{black}{
    (i.) From version 2.5 of HLF, the transaction flow begins with the client application sending a transaction proposal to a \textit{gateway}, which acts as an intermediary between the client and the peers;
    (ii.) Endorsing peers will simulate the transactions and determine if they meet the pre-established requirements in the application's smart contract, which is a self-executing contract with business rules directly written into code;
    (iii.) The transaction is then forwarded to the ordering peers (\textit{orderer}), who will insert it into a block together with other transactions;
    (iv.) The generated block is subsequently persisted on the blockchain through the \textit{commit} process.
}

\subsection{Proposed Model}

\textcolor{black}{Stochastic modeling is a critical tool for assessing blockchain performance\cite{Melo2022, Sukhwani2018}. Melo et al.\cite{Melo2022} introduced models for evaluating Hyperledger Fabric applications' availability and costs. Sukhwani et al.\cite{Sukhwani2018} used Stochastic Reward Nets (SRN) to analyze throughput and utilization. This approach uses probabilistic models to simulate and evaluate system behavior under various conditions\cite{Maciel2012}. By applying random variables and statistical distributions, stochastic modeling provides insights into blockchain performance metrics\cite{Melo2022, Sukhwani2018}.
}

\blue{We propose a Stochastic Petri Net (SPN)} to model the Fabric environment, as shown in Figure \ref{fig_spn_model}. This tool effectively represents concurrent systems\cite{Maciel2012}.
The transaction flow initiates at the \textbf{Arrival} transition, \blue{where tokens are generated; these tokens represent a live transaction proposed by a client and arrive in the system at a predefined arrival rate.}
\blue{Each transaction that reaches the system is directed to \textbf{P0} place that represents an endorsement node or set of nodes, which contains an endorsement queue (\textit{en\_q}) associated with it signifying that we have limited resources. Post-endorsement, transactions move to the \textbf{P1} place at a pre-established endorsement time (Endorse transition)}, waiting to be ordered into a block.
\blue{An endorsed transaction waits to be ordered by the firing of the Order transition, which has the predefined ordering time associated with it. An ordered transaction advances to the \textbf{P2}, which correlates to the \textbf{blo\_q} place, that defines the block size. This size is adjustable according to specific needs. Upon reaching the block capacity, defined by \textbf{q\_size}, the block of transactions moves from \textbf{P2} place to the \textbf{block} place via the \textbf{FullBlock} transition, forming a block that is then committed to the network. Alternatively, blocks of transactions can move through the \textbf{BatchTimeout} transition if they exceed the set time limit in the ordering phase, preventing transaction delays.}
The \textbf{flush} variable indicates that the \textbf{P2} place will be cleared once the \textbf{BatchTimeout} transition activates, ensuring all tokens from \textbf{P2} are moved to the \textbf{blo\_q} place without exceeding \textbf{q\_size}.
Guard expressions are integrated to control the activation of specific transitions: \textbf{FullBlock} and \textbf{BatchTimeout} to enhance system precision. The \textbf{BatchTimeout} transition fires only when there is at least one transaction in the block, while \textbf{FullBlock} activates when the block reaches total capacity.

\textcolor{black}{The model in \cite{Melo2022} by Melo et al. uses Stochastic Petri Nets (SPNs) to assess resource utilization in Hyperledger Fabric, focusing on resource allocation and sensitivity analysis for detecting infrastructure bottlenecks. In contrast, our model targets key performance metrics such as mean response time (MRT), throughput, and resource utilization. We provide a detailed breakdown of the transaction flow within Hyperledger Fabric, including the endorsement, ordering, and committing phases, and integrate formal model validation with sensitivity analysis for bottleneck detection. Additionally, our model includes extensive experimental validation to ensure accuracy in simulating real-world scenarios, which is not comprehensively addressed in \cite{Melo2022}.}

\section{Experimental Methodology}
\label{sec:measurement}

\textcolor{black}{Our model was validated using the Hyperledger Fabric test network, which facilitates asset creation, verification, and transfers. The experimental setup was composed of two peers and one orderer within Docker containers on hardware with four cores, 8GB RAM, and 80GB storage. The software configuration included Ubuntu 22.04, Fabric 2.5, and Docker 24.text}

\textcolor{black}{In this setup, the two peer containers represented organizations, while the orderer container managed transactions, committing them to the blockchain via the network's peers. Communication was facilitated by a client container with a command-line interface (CLI). The TypeScript client application managed transaction submissions at 20 per second, validating the model's performance. Measurements were performed in this environment with 100 simulations per model parameter, totaling nearly 500 transactions. The Mercury Modeling Tool \cite{Maciel2017} calculated the mean response time (MRT), yielding an experimental MRT of 1334ms (standard deviation 194ms) and a simulated MRT of 1320ms (standard deviation 206ms). A two-sample T-test with a p-value of 0.78 at a significance level of 0.05 revealed no significant differences between experimental and simulated data, indicating high model accuracy in replicating system dynamics. The confidence level was 95\%.}
\textcolor{black}{We acknowledge the importance of comparative experiments to demonstrate the effectiveness of our method. However, direct comparisons with existing studies are challenging due to differences in Hyperledger Fabric versions and the lack of comprehensive validation and bottleneck detection in prior research. Most studies focus on older versions and theoretical models without extensive experimental validation. Our work addresses these gaps by introducing a robust Stochastic Petri Net (SPN) model tailored for HLF v2.5+, validated through rigorous experiments. We evaluate key performance metrics such as mean response time (MRT), throughput, and resource utilization, providing detailed sensitivity analysis and formal bottleneck detection. These advancements ensure our contributions are relevant and applicable to current and future HLF deployments, offering significant value beyond the scope of previous research.}


\section{Case Studies}
\label{sec:results}

\textcolor{black}{The validated model was applied to scenarios to assess blockchain metrics: mean response time, throughput, and utilization. Sensitivity analysis calculated percentage differences to measure component impacts \cite{Greenland2024}. Baseline values, from average results, are outlined in Table \ref{tab_variation} and serve as input parameters for the model in our case studies.}

\begin{table}[htpb]
\caption{Factor variation and input values}
\centering
\footnotesize
\begin{tabular}{lrr}
\toprule
\multicolumn{1}{c}{\textbf{Factor}} & \multicolumn{1}{c}{\textbf{Baseline Value}} & \multicolumn{1}{l}{\textbf{Variation Range}} \\ \hline
Arrival Rate         & 10/s  & \{7, 18\} \\
Block Timeout        & 2000ms & \{1000, 3000\} \\
Committing Time      & 1150ms & \{575, 1725\}  \\
Ordering Time        & 15ms   & \{7.5, 22.5\}  \\
Endorsing Time       & 160ms  & \{80, 240\}    \\
Block Size           & 10     & \{5, 45\}      \\
Endorsing Queue Size & 10     & \{5, 15\}      \\ 
\bottomrule
\end{tabular}
\label{tab_variation}
\end{table}

\subsection{Case Study I - mean response time (MRT)}

The first metric is the MRT, which can be derived from Little's Law\cite{jainart}. Little's Law establishes a relationship between the average number of transactions in progress within a system and the transaction arrival rate, as expressed in Equation \ref{eq_general_mrt}.

\vspace{-10pt}
\begin{equation}
\label{eq_general_mrt}
\textbf{MRT} = \frac{\sum_{i} E(\text{Place}_i)}{\text{Arrival Rate}}
\end{equation}
\vspace{-10pt}

In the context of stochastic models, such as the proposed Stochastic Petri Net (SPN), calculating the number of transactions in progress within the system involves summing the number of tokens or resources in each place representing a transaction in progress. 
In this context, $E(\text{Place})$ denotes the statistical expectation of tokens in a given place, as present in Equation \ref{eq_expected}.

\vspace{-10pt}
\begin{equation}
\label{eq_expected}
\textbf{E(\text{Place})} = (\sum_{i=1}^{n} P( m(\text{Place})=i)\times i)
\end{equation}
\vspace{-10pt}

where \textbf{P(m(\text{Place})=i)} stands for the probability that there are \textit{i} tokens or transactions in the given place.

\begin{figure*}[htpb]
    \centering
    
    \begin{subfigure}[b]{0.32\textwidth}
        \includegraphics[width=\textwidth]{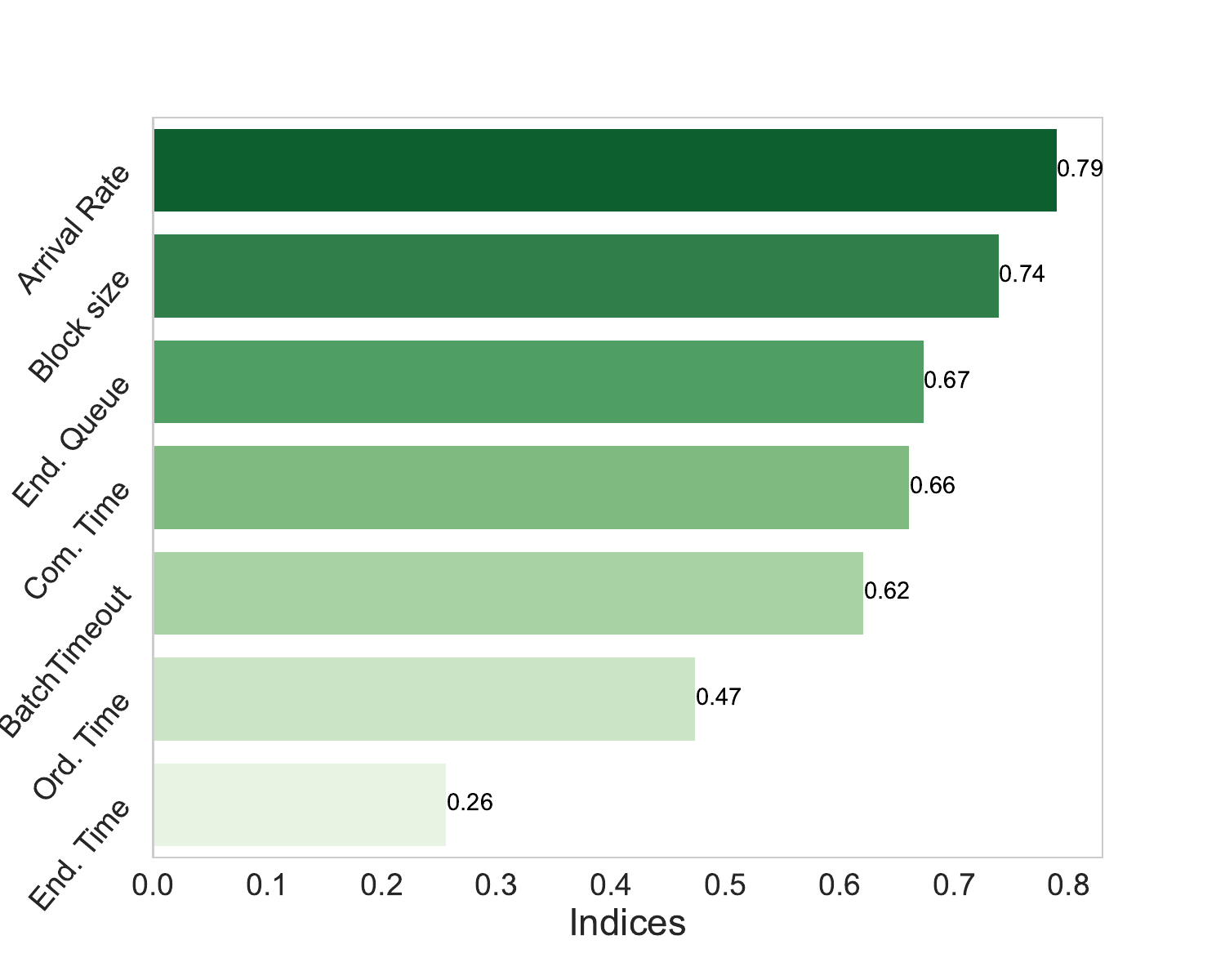}
        \caption{Sensitivity Indices for MRT}
        \label{fig_sensitivity_mrt}
    \end{subfigure}
    \hfill
    \begin{subfigure}[b]{0.32\textwidth}
        \includegraphics[width=\textwidth]{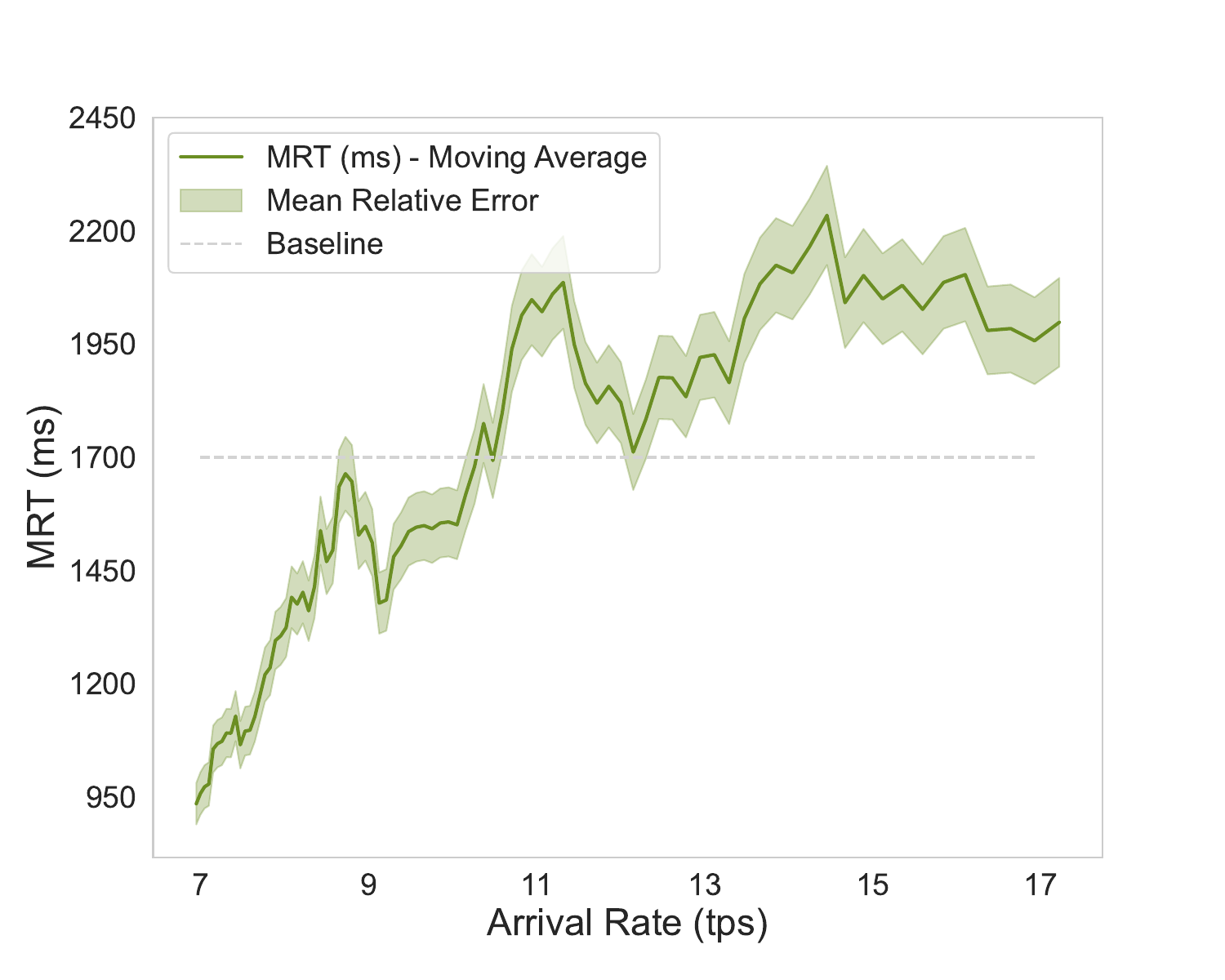}
        \caption{Arrival Rate}
        \label{fig_arrival_mrt}
    \end{subfigure}
    \hfill
    \begin{subfigure}[b]{0.32\textwidth}
        \includegraphics[width=\textwidth]{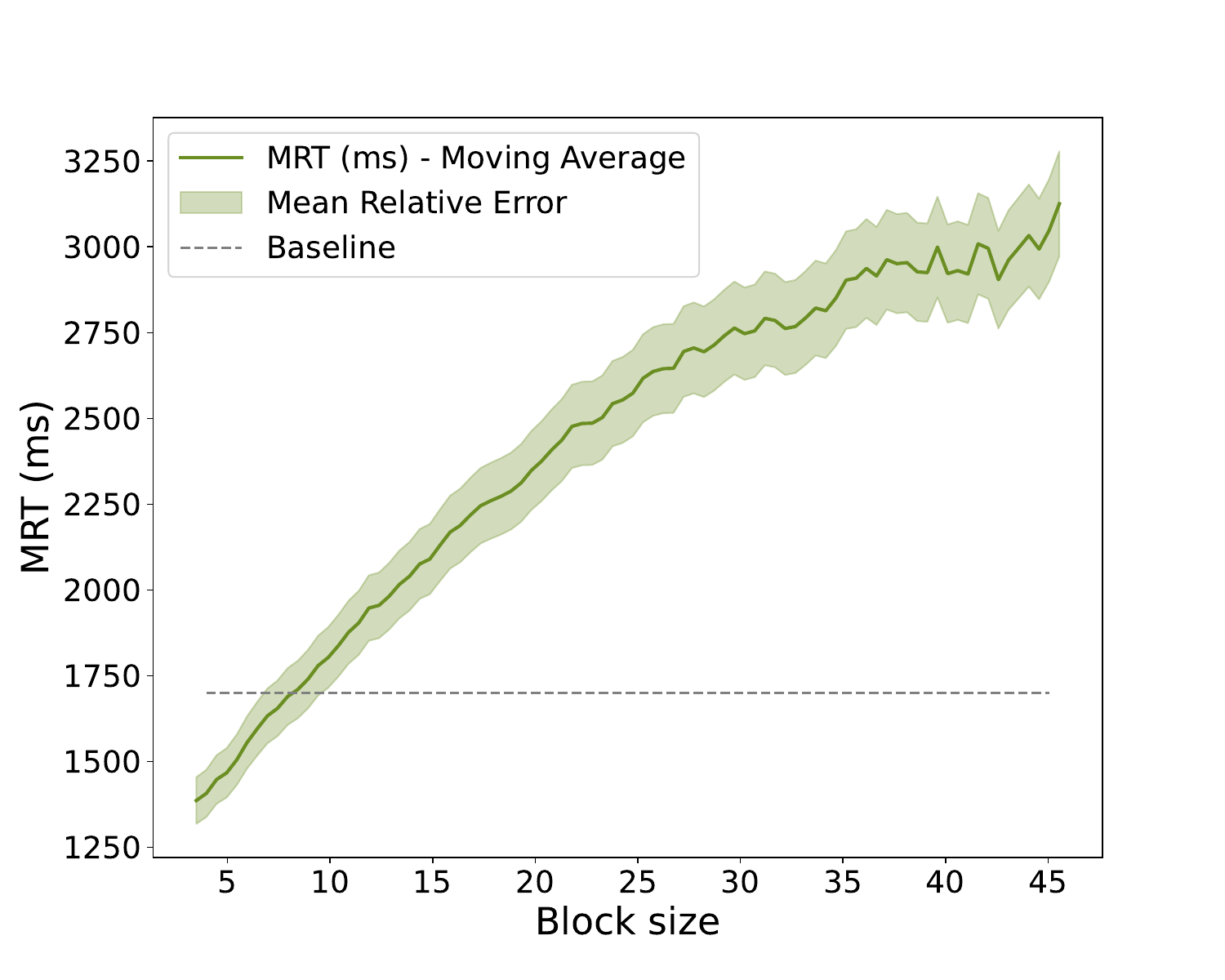}
        \caption{Block size}
        \label{fig_block_mrt}
    \end{subfigure}
    \vspace{-10pt}
    \caption{Sensitivity Analyses of MRT wrt. Impacting Factors}
    \label{fig_mrt_analysis}
    \end{figure*}




\textcolor{black}{The baseline scenario evaluation showed a mean response time (MRT) of 1.754 ms. MRT's sensitivity to various components was analyzed, with percentage impact factors from 0\% to 100\%. Figure \ref{fig_sensitivity_mrt} reveals the primary influences: \textit{Com. Time}, \textit{Ord. Time}, \textit{End. Time}, and \textit{End. Queue} (Committing Time, Ordering Time, Endorsing Time, and Endorsing Queue Size, respectively). Arrival Rate impacts MRT significantly at 79\%, followed by Block Size. Figure \ref{fig_arrival_mrt} shows the arrival rate's effect on MRT, indicating higher transaction numbers increase MRT, potentially overloading the system. Figure \ref{fig_block_mrt} illustrates how smaller block sizes increase MRT due to more frequent block creation and dispatching to the commit phase.}

\subsection{Case Study II - throughput (TPS)}

\textcolor{black}{The throughput metric counts transactions performed by the system over a specified timeframe \cite{thakkar2018performance} and relates to a given transition and place in stochastic models. This paper focuses on committed transactions persisted in the blockchain. Equation \ref{eq_t} calculates system throughput, with \textit{t} representing the timed transition.xt}

\vspace{-10pt}
\begin{equation}\label{eq_t}
	\begin{aligned}
		\textbf{Throughput} = \frac{\text{E(Place)}}{\text{t(Transition)}}
	\end{aligned}
\end{equation}
\vspace{-10pt}

For the proposed model, \textbf{E(Place)} refers to the expected amount of transactions in \textbf{block} and \textbf{P2} places, which is an estimation since a set of transactions can be merged into a single block, and a block does not always have the same size as another (BatchTimeout).



\begin{figure*}[htpb]
    \centering
    \begin{subfigure}{0.32\textwidth}
        \includegraphics[width=\linewidth]{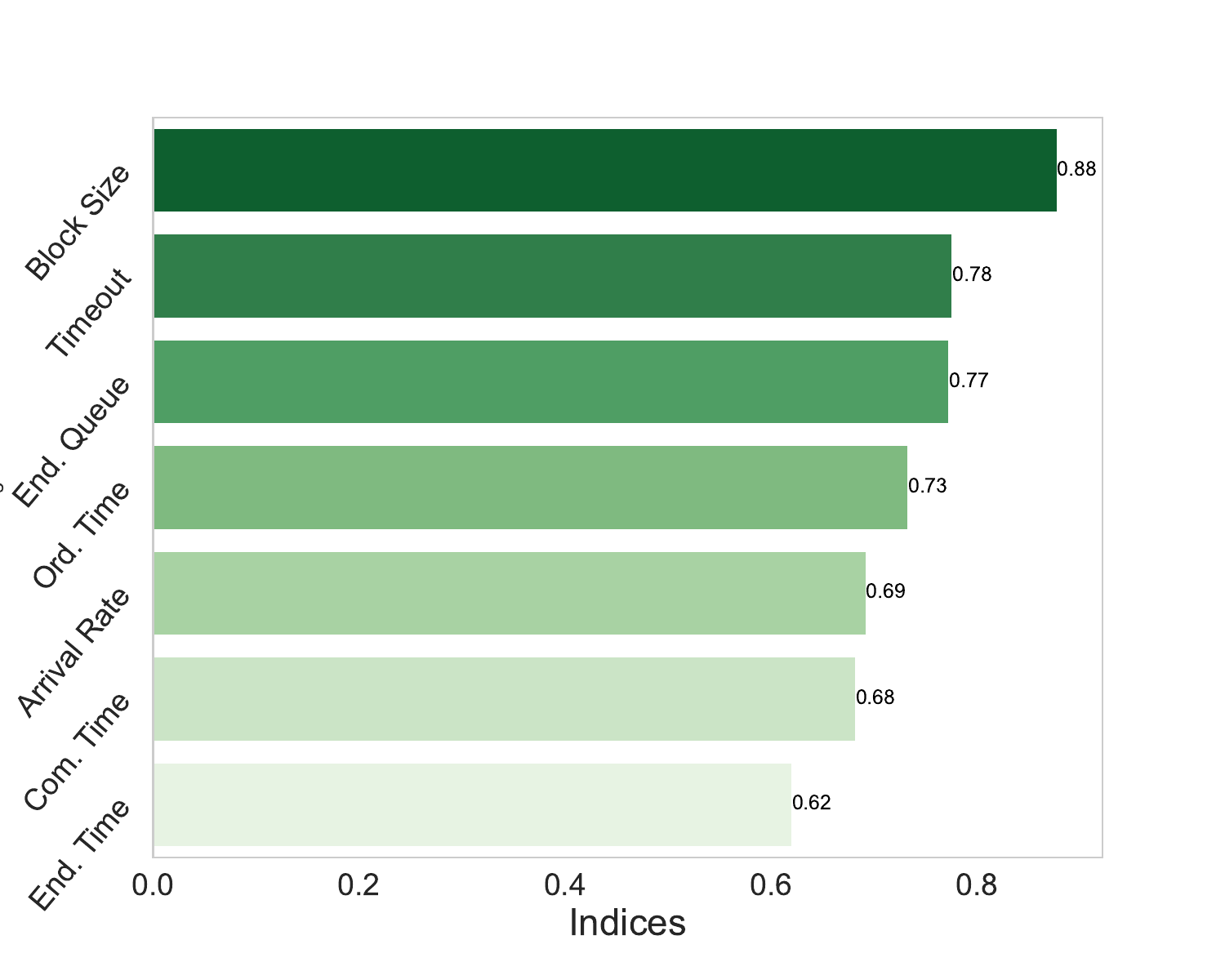}
        \caption{Sensitivity Indices for throughput}
        \label{fig_sensitivity_tps}
    \end{subfigure}
    \begin{subfigure}{0.32\textwidth}
        \includegraphics[width=\linewidth]{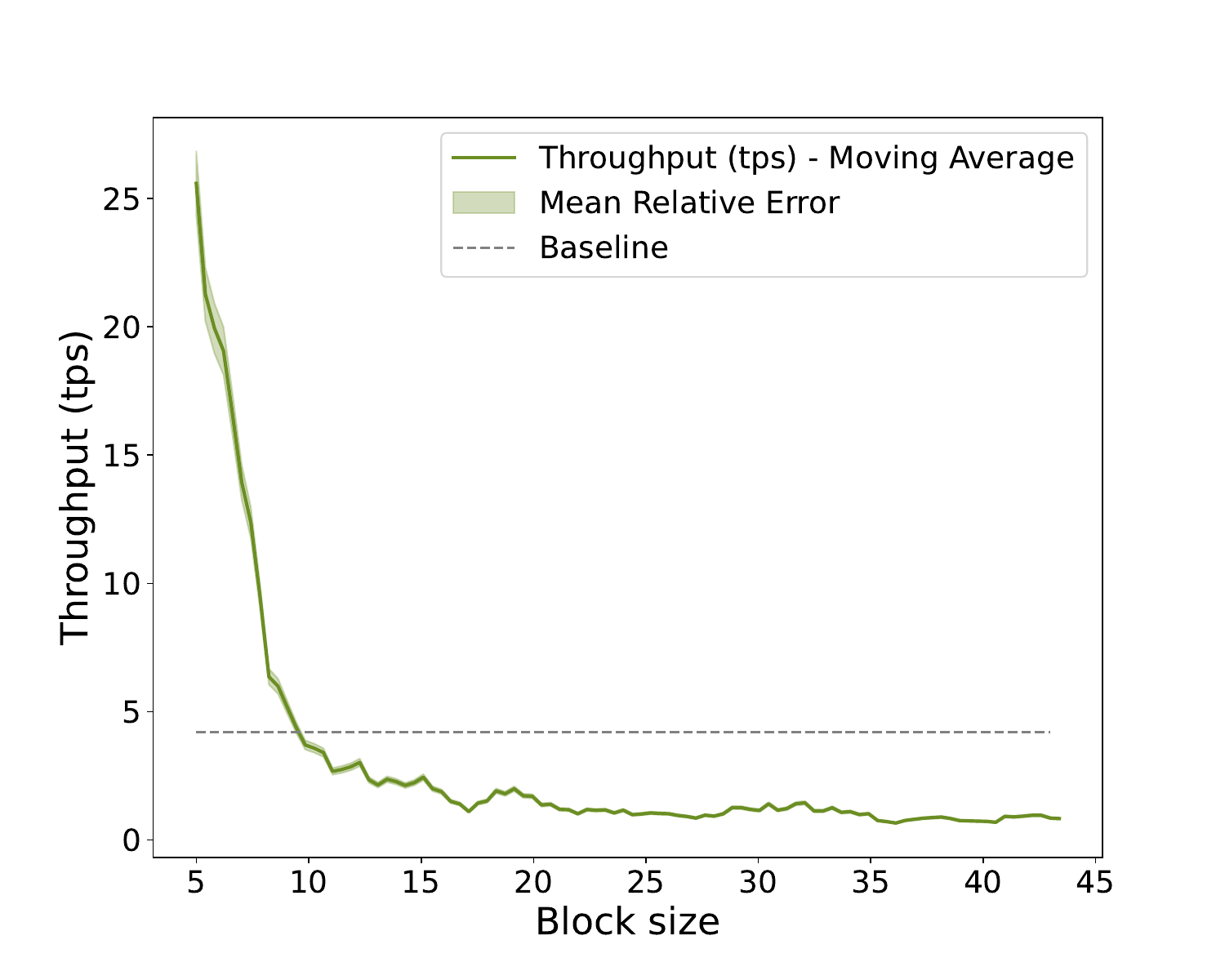}
        \caption{Block Size}
        \label{fig_block_tps}
    \end{subfigure}
    \begin{subfigure}{0.32\textwidth}
        \includegraphics[width=\linewidth]{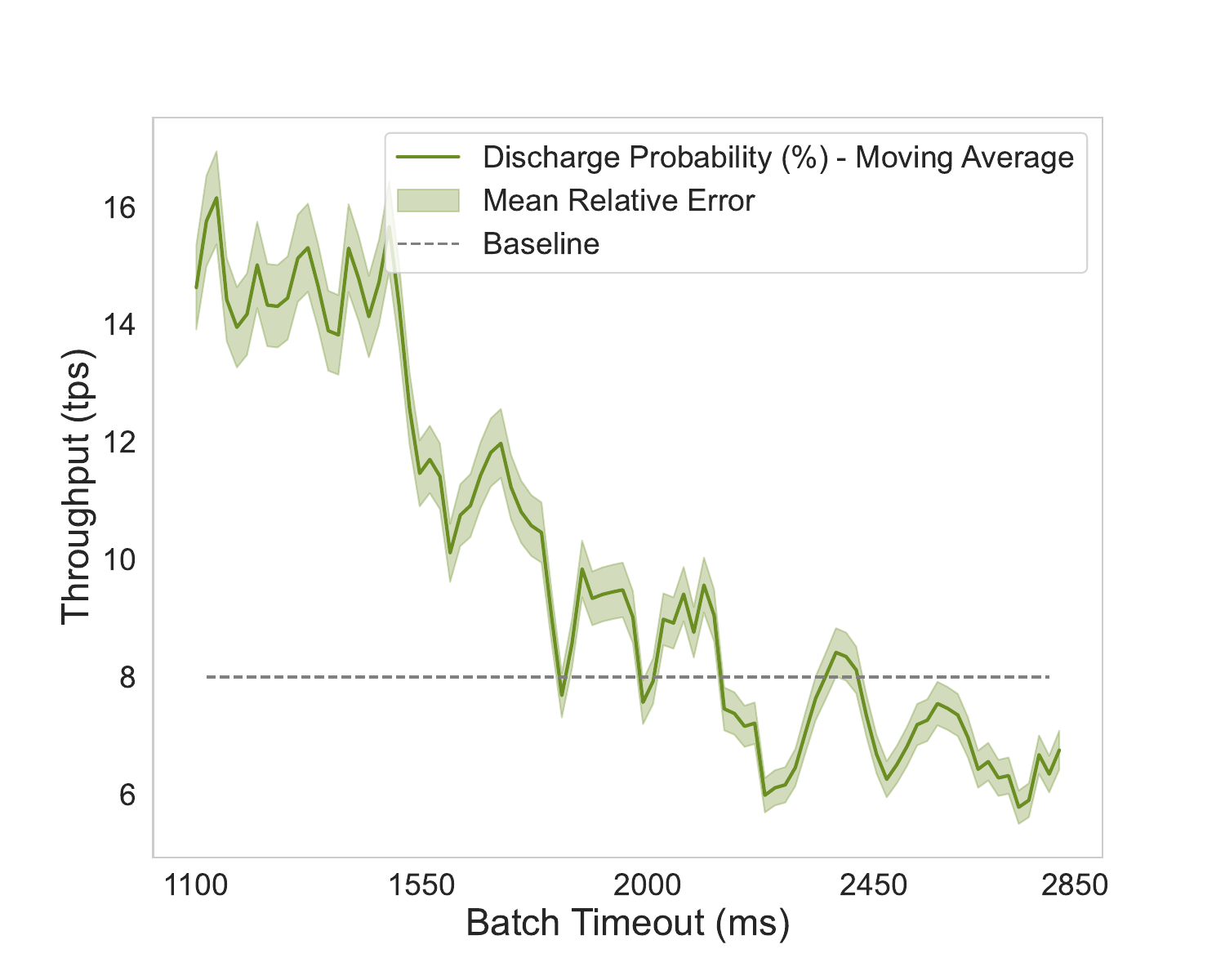}
        \caption{Batch Timeout}
        \label{fig_batch_tps}
    \end{subfigure}
    \vspace{-10pt}
    \caption{Sensitivity Analyses of Throughput wrt. Impacting Factors}
\end{figure*}

\textcolor{black}{The baseline scenario evaluation showed a throughput of nearly eight transactions per second. During sensitivity analysis, Figure \ref{fig_sensitivity_tps} identifies the primary components influencing this value. Block size has the most significant impact, serving as the primary bottleneck and inversely affecting mean response time (MRT) and throughput. Figure \ref{fig_block_tps} illustrates that smaller block sizes increase throughput due to quicker transaction flow through endorsement, ordering, and committing phases. When the block size reaches 15, throughput drops to less than one transaction per second, with batch timeout becoming the primary factor in block generation. Figure \ref{fig_batch_tps} shows that increased batch timeout reduces TPS, leading to longer transaction wait times and larger block sizes.}

\subsection{Case Study III - Utilization}

\begin{figure*}[htpb]
\centering
\begin{subfigure}{0.32\textwidth}
  \includegraphics[width=\linewidth]{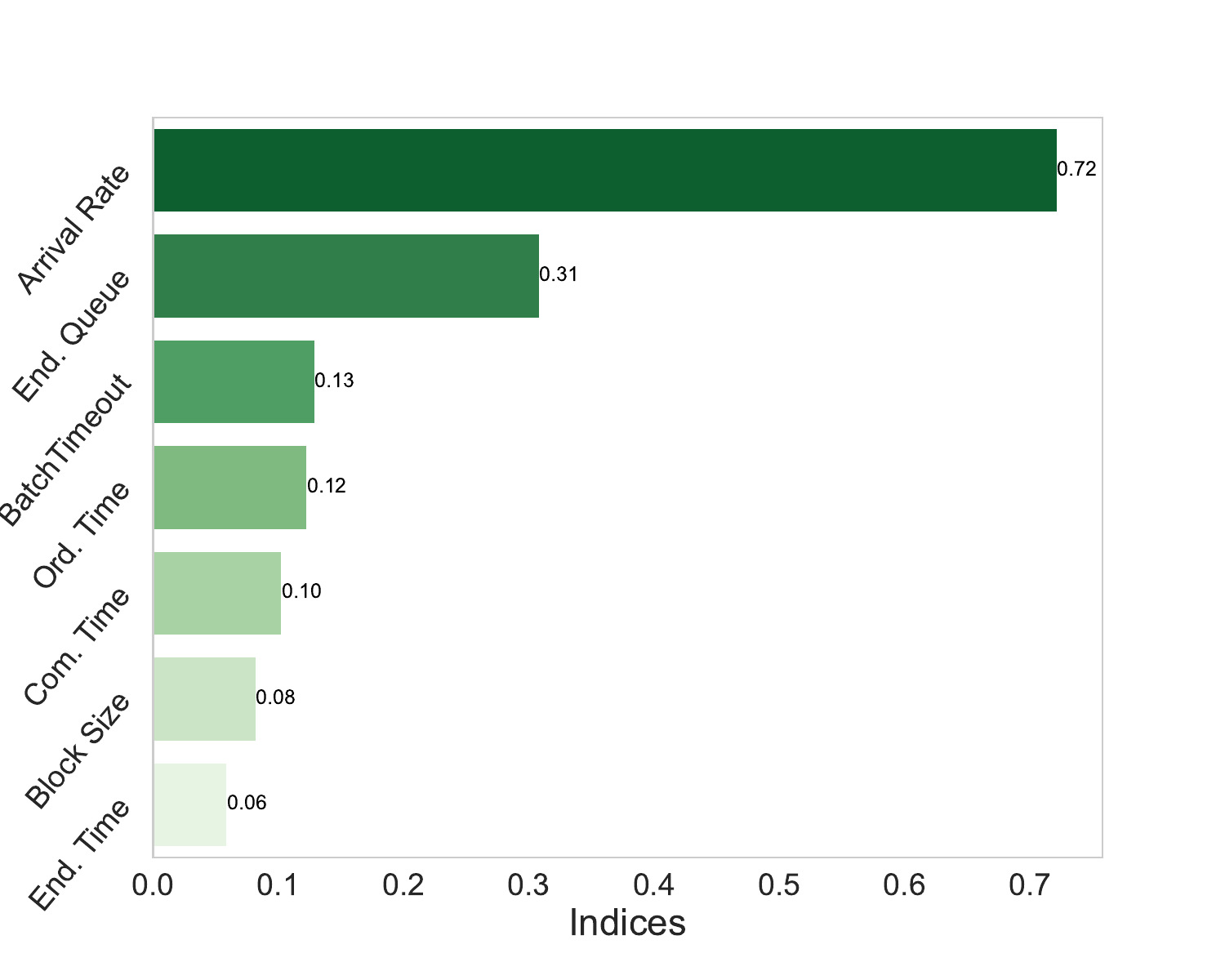}
  \caption{Sensitivity Indices for Utilization}
  \label{fig_sensitivity_utilization}
\end{subfigure}
\begin{subfigure}[b]{0.32\textwidth}
    \includegraphics[width=\textwidth]{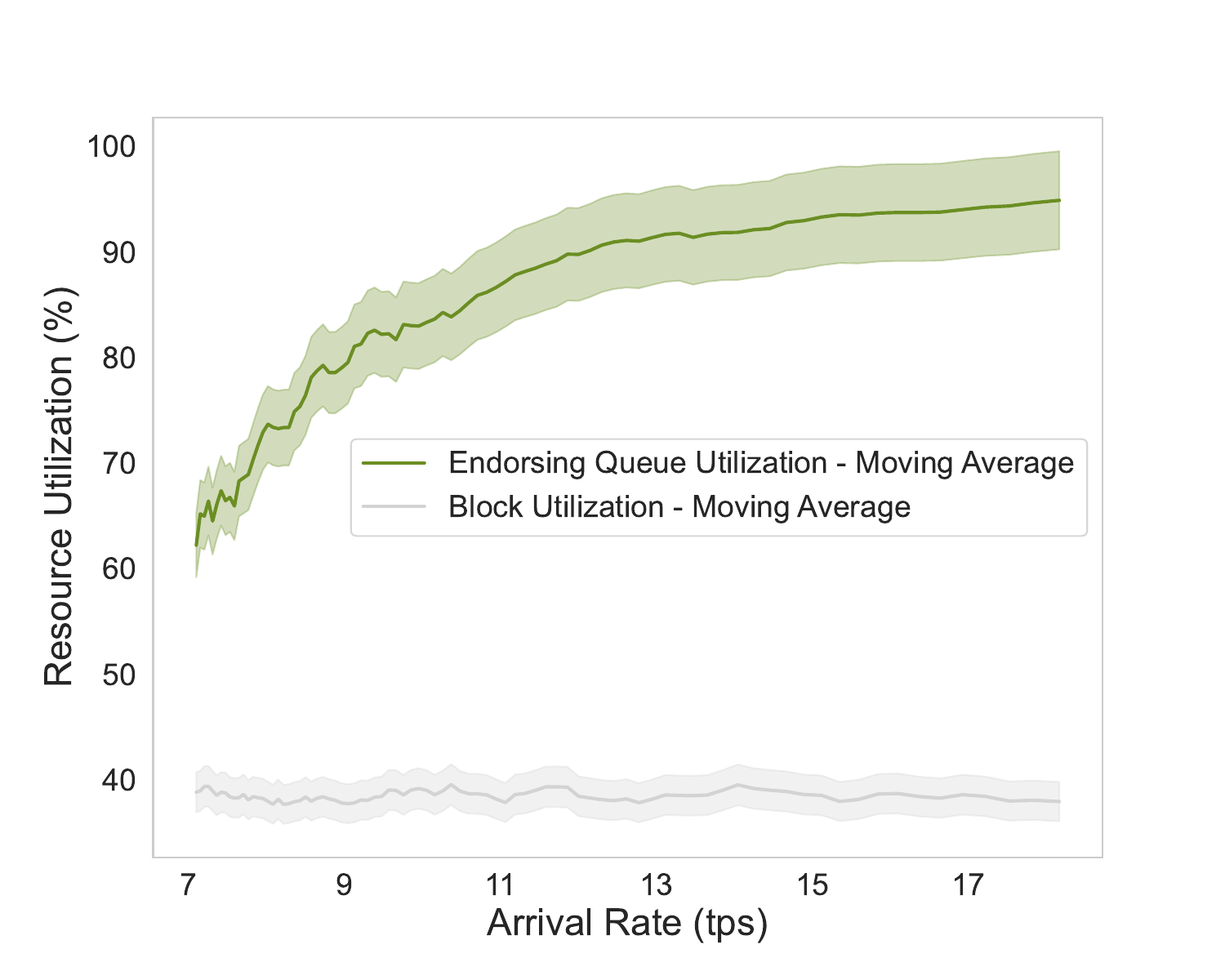}
    \caption{Resource Utilization}
    \label{fig_utilization}
\end{subfigure}
\begin{subfigure}[b]{0.32\textwidth}
    \includegraphics[width=\textwidth]{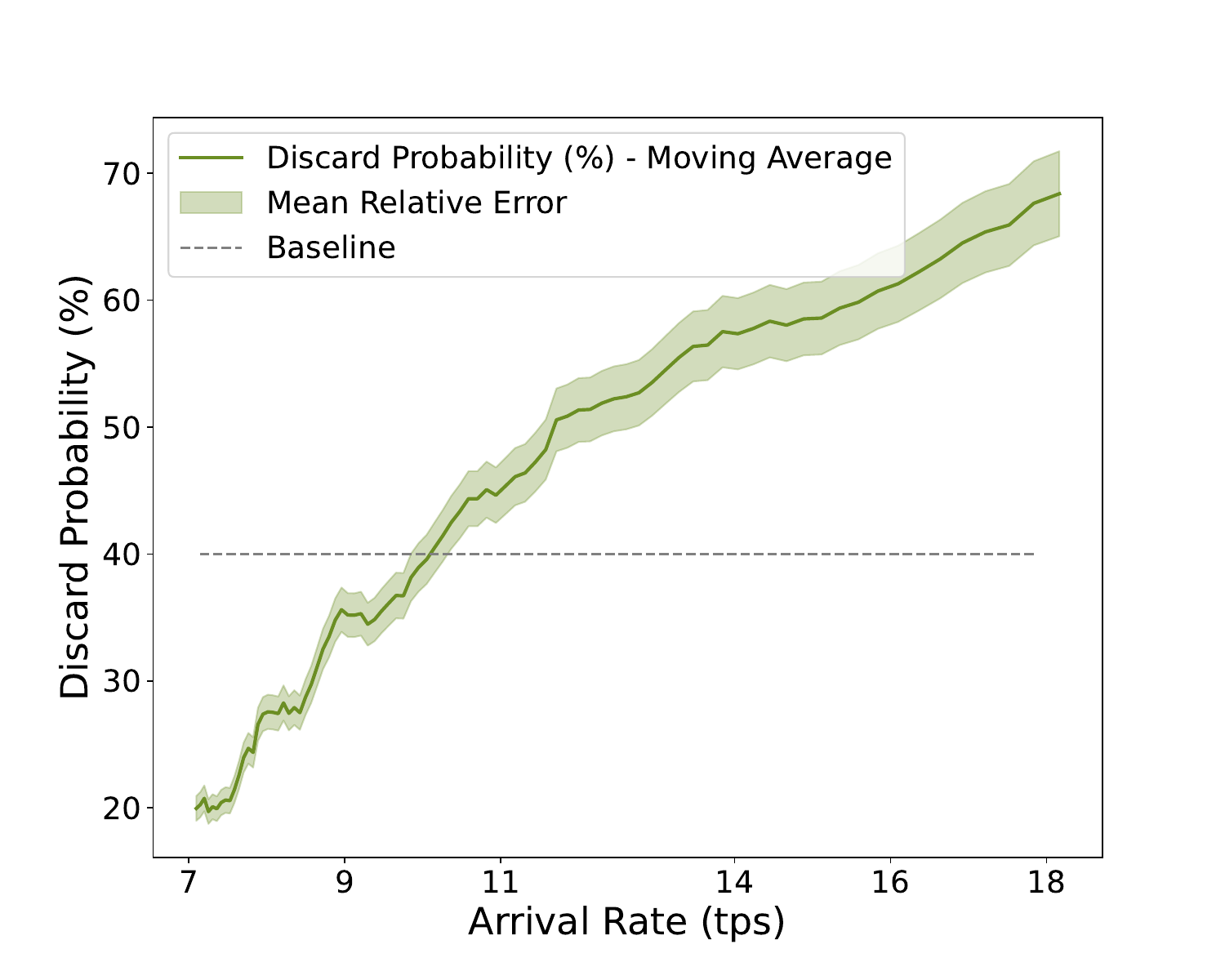}
    \caption{Discard Probability}
    \label{fig_dischard}
\end{subfigure}
\vspace{-10pt}
\caption{Utilization Analysis}
\label{fig_utilizations}
\end{figure*}

When dealing with SPNs, the resource utilization quantifies the expected number of resources in a particular place (where tokens traverse)\cite{Maciel2012}, divided by the total capacity of the directed related place, effectively characterizing it as a queue, as shown in Equation \ref{eq_u}.

\vspace{-10pt}
\begin{equation}\label{eq_u}
\begin{aligned}
\textbf{Utilization} = \frac{\text{E(Place)}}{\text{Resources Capacity}}
\end{aligned}
\end{equation}
\vspace{-10pt}

\textcolor{black}{The proposed model quantifies utilization at two critical points: the endorsement queue (\textbf{en\_q}), where transactions await peer simulation and approval, and the block queue (\textbf{blo\_q}), where transactions are assembled into blocks for commitment. Block size significantly influences performance indicators, including mean response time (MRT) and throughput.}
\textcolor{black}{Sensitivity analysis on utilization, depicted in Figure \ref{fig_sensitivity_utilization}, reveals that arrival rate significantly impacts utilization. Figure \ref{fig_utilization} shows average utilization for both queues, with the endorsement queue often at full capacity (100\%), assuming a 5\% error margin. Higher arrival rates degrade performance due to limited queue capacity. Throughput is constrained by block queue size, while arrival rate primarily affects MRT. The ordering phase is approximately 10\% of the endorsing phase, which is 10\% of the committing phase. Optimizing system capacity is essential; large blocks reduce MRT but stabilize throughput due to batch timeout, while smaller blocks enhance throughput by completing faster. Minimal endorsement times are crucial for reducing MRT and boosting throughput.}
\textcolor{black}{High arrival rates can strain the system, increasing transaction discards at the endorsement stage. The transaction discard probability was analyzed to indicate when the \textbf{en\_q} reaches zero capacity, preventing new transactions. This probability, relative to arrival rate, is illustrated in Figure \ref{fig_dischard}.}

\section{Conclusion}
\label{sec:conclusion}

\textcolor{black}{This paper presents a detailed performance analysis of Hyperledger Fabric, examining bottlenecks in mean response time (MRT), throughput (TPS), and utilization. The model is validated through experiments focusing on the endorsing, ordering, and committing phases, using asset creation transactions from a TypeScript-based client. Exponential time distributions were used, but may need refinement for better scalability.}
\textcolor{black}{The significance of our analysis lies in its ability to provide a comprehensive evaluation of the Hyperledger Fabric (HLF) network's performance using a Stochastic Petri Net (SPN) model. Our approach identifies critical performance bottlenecks in the endorsement, ordering, and committing phases, providing detailed insights into mean response time (MRT), throughput, and resource utilization. By conducting sensitivity analysis, we determine the impact of various system parameters on performance metrics, offering actionable guidance for optimizing system configurations. This analysis helps administrators make informed decisions to enhance network efficiency, balance system load, and plan for scalability in real-world HLF deployments.}

\bibliographystyle{cas-model2-names}
\bibliography{references}

\end{document}